\documentstyle[11pt,aaspp4]{article}
\newcommand\beq{\begin{equation}}
\newcommand\eeq{\end{equation}}

\begin{document}

\title{SYNCHROTRON EMISSION FROM HOT ACCRETION FLOWS AND 
THE COSMIC MICROWAVE BACKGROUND ANISOTROPY}

 \author{Rosalba Perna\altaffilmark{1} and Tiziana Di Matteo\altaffilmark{2}}
 \affil{Harvard-Smithsonian Center for Astrophysics,
60 Garden St., Cambridge, MA 02138;\\
rperna, tdimatteo@cfa.harvard.edu}

\altaffiltext{1}{Harvard Junior Fellow}
\altaffiltext{2}{{\em Chandra\/} Fellow}

\begin{abstract} 

Current estimates of number counts of radio sources in the frequency
range where the most sensitive Cosmic Microwave Background (CMB)
experiments are carried out significantly under-represent sources with
strongly inverted spectra.  Hot accretion flows around supermassive
black holes in the nuclei of nearby galaxies are expected to produce
inverted radio spectra by thermal synchrotron emission.  We calculate
the temperature fluctuations and power spectra of these sources in the
Planck Surveyor 30 GHz energy channel, where their emission is
expected to peak.  We find that their potential contribution is
generally comparable to the instrumental noise, and approaches the CMB
anisotropy level at small angular scales.  Forthcoming CMB missions, which
will provide a large statistical sample of inverted-spectra sources,
will be crucial for determining the distribution of hot accretion flows
in nearby quiescent galactic nuclei. Detection of these sources in
different frequency channels will help constrain their spectral
characteristics, hence their physical properties.

\end{abstract}


\section{INTRODUCTION}

The upcoming cosmic microwave background (CMB) experiments, e.g.~MAP
and the Planck Surveyor, will be able determine the primordial
anisotropies to an unprecedented level of accuracy. Because of its high
sensitivity, excellent angular resolution and wide range of
frequencies, Planck in particular, will be extremely sensitive to
extragalactic foreground point sources, which provide the major source
of uncertainty in the measurement of the intrinsic fluctuations.

Several studies have therefore been carried out to calculate the
contribution of point sources to the CMB anisotropies.  Much of this
work (see Toffolatti et al. 1999a,b; De Zotti et al. 1999; Gawiser \&
Smoot 1997; Sokasian, Gawiser \& Smoot 1998) has dealt with the
contribution from radio sources, the number counts of which are
determined down to $\mu$Jy but only up to frequencies $\la$ 8 GHz.
These counts are usually extrapolated to the higher frequencies
relevant for the CMB experiments.  This implies that the available
counts are sensitive enough to include the most significant
contribution from the ``steep'' and ``flat'' spectrum sources (with
$F_\nu\propto \nu^{-\alpha}$, and $\alpha\ge 0$, such as compact radio
galaxies and radio loud quasars), but are missing, or are strongly
under-representing, an important contribution from a class of sources
with inverted spectra ($\alpha\la 0$; e.g. De Zotti et al. 1999). This
is further emphasized by recent observations at 28.5 GHz, which find
up to a factor of 7 more sources than predicted from low-frequency
surveys (Cooray et al.~1998). Inverted-spectrum sources, such as those
discussed here, may peak in the frequency range of a few tens to a few
hundreds GHz, and could therefore provide a considerable contribution
in the region where the most sensitive CMB experiments are carried
out.

GHz Peaked Spectrum (GPS) sources (see O'Dea et al.~1998, Guerra,
Haarsma \& Partridge 1998) have been recognized to be an important
class of inverted-spectrum sources.  Their emission is attributed to
synchrotron radiation from compact and high density regions often
associated with the early stages of the formation of more classical
double radio sources (the so called ``young source'' scenario; Philips
\& Mutel 1982). However, as pointed out by Toffolatti et al. (1999),
there may be another, distinct, class of strongly inverted spectra due
to thermal synchrotron emission in hot or advection dominated
accretion flows (ADAFs). Unlike the relatively rare and bright GPS
sources (peak fluxes of $\sim 1-10$ Jy), usually associated with
bright active galaxies or quasars at high redshifts, ADAF sources
should be common in nearby galaxies and provide the most significant
contribution to the emission in the high radio frequencies of the
faint ($\sim$ a few mJy) radio cores observed in such galaxies. 

The reason why we consider hot accretion flows to be common in nearby
galactic nuclei is that, in recent years, it has become apparent
(e.g. Fabian \& Rees 1995; Narayan \& Yi 1995; Di Matteo et al. 2000
and references therein) that the nuclei of such galaxies, which host
the largest black holes known with masses of $10^8-10^{10} M_{\odot}$
(e.g., Magorrian et al. 1998), are remarkably underluminous for the
typically expected accretion rates (determined from measuraments of
densities and sound speeds of their hot interstellar medium). In
particular, it has been shown (e.g. Di Matteo et al. 2000) that the
relative quiescence and spectral characteristics of the early-type
galactic nuclei can be well-explained if the central black holes
accrete via low radiative-efficiency accretion flows or ADAFs (Rees et
al. 1982; for a review see, e.g., Narayan, Mahadevan \& Quataert
1998).  Moreover, it has been proposed (Di Matteo and Allen 1999) that
such flows, which also produce significant emission in the $X$-ray
band, could provide a significant contribution to the cosmic $X$-ray
background (XRB). Within the context of these models, a significant
fraction of the hard number counts in the X-ray energies should arise
from sources at low redshift ($z \la 1$).  This picture is supported
by recent deep {\em Chandra} observations, which have resolved about
40 per cent of the hard XRB in point sources in bright early-type
galaxies (Mushotzky et al. 2000).

The potential contribution of GPS sources to fluctuations in the CMB
anisotropy has been discussed by De Zotti et al. (1999).  In this {\em
Paper}, we examine the specific contribution of inverted spectra ADAF
sources in the nuclei of early-type galaxies to the CMB anisotropy. We
evaluate their foreground contribution to the small-scale cosmic
microwave fluctuations in the low-energy channels foreseen for the
Plank surveyor mission.  These sources, if indeed common in elliptical
galaxies, should be much more numerous albeit fainter than the GPS
population, and may therefore provide a stronger noise contribution at
the small angular scales.

While it is important to assess the potential contribution of
advection-dominated sources to the CMB fluctuations, the forthcoming
CMB experiments themselves will, for the first time, provide a large
statistical sample of objects with inverted radio spectra.  Because
most of the ADAF emission occurs in the high radio and in the $X$-ray
band, Planck observations will possibly provide the most powerful test
for the presence of ADAFs around supermassive black holes.  In
particular, such studies will provide strong constraints on the
spectral properties of this class of objects, and will help determine
how common they are in the nearby Universe. Confirming the presence of
these sources would also support the conjecture that they provide a
significant contribution to the hard XRB.

\section{SYNCHROTRON EMISSION FROM ACCRETION FLOWS IN EARLY-TYPE GALAXIES}

Radio continuum surveys (at $\nu \la $ 8 GHz) of elliptical and S0
galaxies have shown that the sources in radio--quiet galaxies tend to
be extended but with a compact component with relatively flat or
slowly rising radio spectra (with typical spectral indexes of
0.3--0.4). Recent VLA studies at high radio frequencies (up to 43 GHz),
although carried out only on a limited sample of objects, have shown that
all of the observed compact cores have spectra rising up to 
$\sim 20-30$ GHz (e.g., Di Matteo et al.~1999).

Although the low-frequency radio emission in these galaxies might
still have a significant contribution from the scaled-down radio jets
also present in these systems, it has been proposed that the
high-frequency emission can be easily accounted for if the supermassive black
holes in elliptical galaxies are accreting via ADAFs (Fabian \& Rees
1995; Mahadevan 1997; Di Matteo et al.~1999). In an ADAF around a
supermassive black hole, the majority of the observable emission is in
the high radio and X--ray bands.  In the high-frequency radio band, the
emission results from synchrotron radiation due to the strong magnetic
field in the inner parts of the accretion flow. The X-ray emission is
due either to bremsstrahlung or inverse Compton scattering.  In the
thermal plasma of an ADAF, synchrotron emission rises steeply with
decreasing frequency. Under most circumstances the emission becomes
self--absorbed and gives rise to a black--body spectrum (in the
Rayleigh-Jeans limit) below a turnover frequency $\nu_{\rm c}$. Above
this frequency it decays exponentially as expected from a thermal
plasma, due to the superposition of cyclotron harmonics.

The spectral models and self-consistent temperature profile
calculations for the ADAF flows in the elliptical galaxy cores are
described in detail by Di Matteo et al. (1999; 2000b). Recent
studies have also shown that large outflows may be important in such
low--radiative efficiency accretion flows (Igumenshchev \& Abramowicz
1999; Blandford \& Begelman 1999; Stone, Pringle \& Begelman 1999). This
should lead to a suppression of the synchrotron component with
respect to the standard ADAF model with no outflow. Flatter density
profiles, as expected from strong mass loss, are usually required to explain the 
high-resolution VLA observations at high-radio frequencies of a number of
ellipticals. As in our previous work, therefore, we model the accretion
flows by adopting a density profile which satisfies $\rho \propto
R^{-3/2 +p}$, where $R$ is the radius of the flow and $0 \le p
\le 1$. Figure 1 shows the synchrotron emission expected from an ADAF
around a supermassive black hole with $M_{BH} \sim 10^9 M_\odot$, for
different values of $p$ (effectively for radially dependent mass
accretion rates, $\dot{M} \propto R^p$). The uppermost curve is the
standard ADAF model calculated with the accretion rates in the flows
determined from Bondi accretion theory, $\dot{M} \sim \dot{M}_{\rm
Bondi} \propto M_{\rm BH}^2 n$(ISM)$/c_{\rm S}$(ISM); where the ISM
density, $n$(ISM), and sound speed, $c_{\rm S}$(ISM), can be
determined from X-ray deprojection analysis, and $M_{\rm BH}$ is given
in recent studies by Magorrian et al. (1998).

So far, only a limited sample of sources has been observed at
high--resolution radio frequencies (up to 43 GHz), such that both the
flux and the position of the synchrotron peak could be determined. The
shaded region in Figure 1 identifies the range of models that have
been shown to best fit the observed fluxes. However, as the sample is
not statistically significant, we will also consider various
populations of accretion sources which may have different luminosities
due to their different density profiles.  Note also, from
Figure 1, that because of their sharp spectral cut-offs, these sources
are expected to contribute only to the Planck low-frequency channels
($\nu \la 100$ GHz), and in particular to the 30 GHz one.

It is worth pointing out that the synchrotron emission, and the
position of its peak in an ADAF, is a strong function of many model
variables: the emission at the self-absorbed synchrotron peak arises
from the inner regions of an accretion flow and scales as $L_{\nu}
\propto \nu_{c}^2 T$, where $\nu_{c} \propto T^{2} B \propto T^2
\dot{M}^{1/2} M_{\rm BH}^{1/4} R^{-5/4}$ (Narayan \& Yi 1995); $T$ is the electron
temperature, and $B$ the magnetic field strength.  Because of the
strong dependences on a number of parameters, we will estimate the
contribution to the CMB anisotropy by making the most conservative
assumptions for the model source parameters.  We stress that our
analysis is not intended to explore the full range of parameter space
available for the accretion flow models. At present, the theoretical
and observational uncertainties involved in such calculations are too
large to merit such work. However, although schematic, these models
may provide a useful guide for the prediction of the confusion noise
due to these sources.

\section{CONTRIBUTION TO CMB FLUCTUATIONS}

The contribution to CMB fluctuations from randomly distributed sources
has been extensively discussed in the literature (Scheuer 1957, 1974;
Condon 1974; Cavaliere \& Setti 1976; Franceschini et al. 1989;
Tegmark \& Efstathiou 1996).

We let $x=Sf(\theta)$  be the telescope response to a source of flux $S$
located at a distance $\theta$ from the beam axis, and let
$R(x)$ be the mean number of source responses of intensity $x$. 
The fluctuation level generated by randomly distributed sources with
a Poisson distribution is given by the second moment $\sigma$ of the
$R(x)$ distribution. If the angular power pattern of the detector,
$f(\theta)$, is taken to be a Gaussian with full width half maximum 
$\theta_0$, one obtains:
\beq
\sigma^2=\int_0^{x_c}x^2 R(x) dx=\pi\theta_0^2 I(x_c)\;
\label{eq:sigma}
\eeq 
where
\beq
I(x_c)=\int_0^{x_c}dx x^2\int_0^{\infty}d\psi N\left(\frac{x}{f(\psi)} \right)
\exp(4\psi\, {\rm ln}2)\;. 
\label{eq:Ic}
\eeq Here $\psi\equiv(\theta/\theta_0)$, $N(S_\nu)$ are the differential
source counts per steradian at a given frequency $\nu$, and $x_c=q\sigma$
is the threshold flux above which sources are considered to be
individually detected.  In our calculations we adopt the standard value of $q=5$.

The rms brightness temperature fluctuation $(\Delta T/T)_{\rm rms}\equiv
\left< (\Delta T/T)^2\right>^{1/2}$ at the  frequency $\nu$ is related to 
the confusion standard deviation $\sigma$ by
\beq
\left(\frac{\Delta T}{T}\right)_{\rm rms}=\frac{\sigma}{T\omega_b}
\left(\frac{\partial B_\nu}{\partial T}\right)^{-1}\;,
\label{eq:delT}
\eeq
where $\omega_b=\pi\int_0^\infty d\theta^2 f(\theta)=\pi(\theta_0/2)^2\log 2$
is the effective beam area, and 
\beq
\frac{\partial B_\nu}{\partial T} = \frac{2k}{c^2}\left(\frac{k T}{h}\right)^2
\frac{x^4 e^x}{(e^x-1)^2} 
\label{eq:conv}
\eeq 
is the conversion factor from temperature to flux (per steradian).  
Here $B_\nu (T)$ is the Planck function, $x\equiv h\nu/kT$,  
and $T=2.725$ K (Mather et al. 1999) is the CMB temperature. 

We take the number density of our sources to be that of ellipticals, and 
use the fit\footnote{The fit was derived from   
observations of over 1700 galaxies in various magnitude
limited samples (the Autofib Redshift Survey).} 
provided by Heyl et al. (1997) for galaxies with $L\approx L_*$,
\beq
n(z) = n_0 (1+z)^{[\gamma_\phi-\gamma_L(\alpha_0+\gamma_\alpha z)]}
\exp[-1/(1+z)^{\gamma_L}]\;,
\label{eq:nz}
\eeq where the set of parameters \{$n_0,\gamma_\phi,\gamma_L,\alpha_0,
\gamma_\alpha $\} is given by Heyl et al. for red and blue
ellipticals. These are \{$1.62\times 10^{-3}$ Mpc$^{-3}$, -6.15,
-1.77, -1.05, 1.3\} and {$1.87\times 10^{-3}$ Mpc$^{-3}$, 1.56, -0.35,
-1.06, 1.23\} for red and blue respectively. We consider the total
$n(z)$ and add both components. We take a cutoff at $z\sim 1$ for
their contribution. This is consistent with the normalization of the
ADAF sources in ellipticals required to explain the hard XRB (Di
Matteo \& Allen 2000). Notice, however, that our results are rather
insensitive to the precise value of this cut-off, as most of the
contribution is produced by sources at very low
redshift.

Given a population of low-redshift sources with luminosity per unit
frequency $L_\nu$, the differential number counts are given by
$N(S_\nu)\approx [4\pi(c/H_0)n(z)d_L^2(z)dz/dS_\nu]_{z=z(S_\nu)}$,
where $z(S_\nu)$ is derived by inverting the relation $S_\nu=L_{\nu}
(1 + z)/4\pi d_L^2(z)$, and $d_L$ is the luminosity distance. We adopt
a flat cosmology with $\Omega_0=1$, and take $H_0=3.2\times 10^{-17}
{\rm s}^{-1} h$, with $h=0.65$. The particular choice of cosmology is
not relevant to our calculations, again because the bulk of the
contribution from these relatively faint objects comes from nearby
sources.

For a Poisson distribution of point sources, it is 
straightforward to compute their angular power spectrum $C_l$.
If only the contribution  from sources with $S_\nu\le S_{\rm lim}$
is included, this is given by (Tegmark \& Efstathiou 1996; Scott \& White 1999)
\beq
C_l(\nu)= \frac{1}{T^2}\left(\frac{\partial B_\nu}{\partial T}\right)^{-2}
\int_0^{\rm S_{\rm lim}} S_\nu^2\; N(S_\nu) dS_\nu\;. 
\label{eq:cl}
\eeq
~~~
~~~

\section{RESULTS}

Figure 2 shows the expected temperature fluctuations (Eq.~3) due to
synchrotron emission from accretion flows in the nuclei of ellipticals
as a function of the angular scale $\theta_0$, at $\nu = 30$ GHz. The
solid line corresponds to the case where the luminosities of all
sources fall in the region marked by the shaded region of Figure
1. These are ADAF models with a significant amount of outflow and
relatively low luminosities. The other lines show the contribution to
the fluctuations due to a mixed population of sources containing a
fraction $f$ of accretion flows with higher synchrotron
luminosities\footnote{ We cannot attempt to model a proper luminosity
function for these sources, as the observed sample in Di Matteo et
al. (2000b) is too small to allow such modelling.}  (e.g. as expected
from ADAFs with no outflows corresponding to the uppermost curve of
Figure 1). Note that even a small fraction of high-luminosity sources
can have large effects on the level of fluctuations. This is due to
the fact that the source number counts roughly scale as $N(S)\propto f
L^{1.5}$, and therefore the dependence on the luminosity is stronger
than the dependence on $f$. Also note that our derived number counts are 
comparable (if $f=0$) to the number counts extrapolated from low-frequency surveys
by Toffolatti et al. (1999a), but can be up to a factor of 10 higher 
if a considerable fraction of higher-luminosity ($f\ga 0.5$)
sources is considered.

Figure 3 shows, in the case of a fraction $f=0.5$ of high-luminosity
sources in the sample, the expected level of fluctuations if all
sources above a flux $S_{\rm lim}$ were individually identified and
substracted out from the sample. This is shown for different choices
of $S_{\rm lim}$. Identification and removal of individual sources can
in principle be done by means of independent surveys.

Using the same parameters as in Figure 3, Figure 4 shows a comparison
between the poissonian power spectrum (Eq.~6) produced by the
accretion flows in ellipticals and the predicted power spectrum for a
standard CDM model. The heavy solid line shows the contribution to the
power spectrum from noise in the 30 GHz channel of the Planck LFI.
Note that the ``flat'' ($C_l$ = constant) angular power spectrum of
the fluctuations due to ADAF sources differs substantially from the
power spectrum of primordial fluctuations. We find that the source signal 
is generally well below that of the intrinsic fluctuations,
and it only becomes comparable to these on the small angular
scales, where also the instrumental noise increases to roughly the
same level.  Removal of source signal should be possible even in cases
where it gives a strong contribution. Even if the Poisson component of
the sources and the noise due to the instrument have similar power
spectra, they are indeed different in their nature (as emphasized by Scott \&
White 1999). In fact, while the sources on the sky contribute to the
flux in every observation of a given pixel, the noise, on the other
hand, differs from observation to observation, and, by assumption, it
is uncorrelated with the signal in that pixel. Therefore, if a given
direction in the sky is observed multiple times (as expected for
Planck), the instrumental noise component can be separated from the
sky signal.

\section{DISCUSSION}

We have computed the temperature fluctuations and power spectrum
produced by inverted radio spectra from hot accretion flows in the
nuclei of nearby elliptical galaxies in the Planck 30 GHz channel,
where their emission is expected to peak. We have shown that the
contribution from this class of sources approaches the intrinsic CMB
fluctuation level only at small angular scales.  However, because of
the different nature of its power spectrum, the source contribution
should not affect the most important goal of the Planck mission, that
is the accurate measurement of the primary CMB anisotropy.

On the other hand, Planck will provide a large statistical sample of
sources characterized by inverted spectra. Therefore, it should be
possible to use this study to determine how common this mode of
accretion is in the nearby supermassive black holes.  In particular,
as most of the contribution from this population is expected to peak
at high radio frequencies, Planck should allow us to study their
spectral characteristics.  In turn, because different spectral
distributions and luminosities reflect the shape of the density
profiles, CMB experiments could allow us to gain important information
on the physical conditions in these accretion flows.  As already noted
by Toffolatti et al. (1999b), the implications of such a study could,
more generally, be significant as a way of testing the physical
processes in the medium surrounding massive black holes, and the
evolution of the interstellar medium in galaxies up to moderate
redshifts. Even more, it would provide a test for current ideas
according to which a significant fraction of the $X$-ray background
may due to accretion in this regime in early-type galaxies in the
local universe (Di Matteo \& Allen 1999). Note that such a significant
statistical study would be more difficult to carry out with surveys at
other wavelengths because of the rapid decline of the ADAF flux, which
makes the emission from this type of accretion flows extremely weak in
the far infrared and optical bands.

We need to stress that, in principle, the contribution from ADAF
sources should be easily disentangled not only from that due to
sources with a flat and steep spectrum, but also from that due to GPS
sources which also have strongly inverted spectra. GPS sources are
typically much brighter (with fluxes typically ranging from a few to
10 Jy) but rarer (usually associated with QSOs) than the expected ADAF
sources. The number of GPS sources rapidly decreases with decreasing
flux, whereas ADAFs are expected to be much more numerous at faint
flux levels.  As a result, GPS are only minor contributors to the
fluctuations at small angular scales, whereas ADAFs would be mostly
significant at these scales.  Therefore it should be possible to study
these two populations independently.

Note that we have shown the expected temperature fluctuations due to
ADAF sources only for the lowest energy channel of Planck.  If most of
the sources are indeed in the range of luminosities consistent with
those observed so far, then this channel is expected to have the
largest (possibly major) contribution, due to the high-frequency
cutoff in the spectrum of these sources. However, if a substantial
population of high-luminosity sources is present, then some
contribution should also be present in the other channels of the
Planck LFI.  The availability of multifrequency data should allow an
efficient identification of pixels contaminated by discrete
sources. In order to carry out a substraction of the contaminating
flux one should therefore take into account that strongly inverted
spectra such as those considered here may not be present in most
frequency channels but give rise to a strong contamination up to a
certain frequency, and then abruptly drop. It should also be pointed
out that, contrary to some of the GPS sources for which variability
has been observed (e.g. Stanghellini et al. 1998), the radio sources in
the hot accretion flows are usually not very variable. A lack of
variability is particularly important for a proper removal of sources
from the spectral fitting.

We note that ADAFs around massive black holes could also be found in
spiral galaxies such as the Galactic nucleus Sgr A$^*$. However, even
if ADAFs were indeed common in spiral nuclei, their potential
contribution to the CMB anisotropy would still be dominated by that
from ellipticals. Inferred black hole masses are found to be
proportional to the mass of the bulge component of their host galaxy,
implying $M_{BH} \sim 10^6-10^7 M_\odot$ for spiral galaxies. As a
consequence, the contribution from spirals should be much lower, as
the radio flux scales as $M_{\rm BH}^{2.5-3}$ (Franceschini et
al. 1998), and (see \S3) their spectrum would peak at frequencies
higher than those of elliptical cores and affect higher energy
(e.g. sub-mm, mm) channels of CMB experiments. Because of this, given
enough sensitivity, the relevance of ADAFs in quiescent spiral nuclei
may also be assesed separately by the forthcoming experiments.

Finally, we note that in our analysis we do not take into account the
effects of source clustering. Clustering decreases the effective
number of objects in randomly placed cells and, consequently, enhances
the cell to cell fluctuations.  There is indeed evidence that the
positions in the sky of a wide variety of extragalactic sources are
correlated (Shaver 1988).  However, the specific correlation function
of our radio--submm sources in early-type galaxies is not
well-constrained. The analyses of Toffolatti et al. (1998) have
shown that the contribution due to clustering (using the two-point
correlation function from sources selected at 5 GHz; Loan, Wall
\& Lahav 1997) is generally small in comparison with the Poisson
term; however, the relative importance of clustering increases if
sources are substracted out from the Planck maps down to faint flux
levels.

\acknowledgements 
We thank Ramesh Narayan for motivating this work,
and Martin White for useful discussions.  We also thank the anonimous
referee whose comments greatly improved the presentation of this
paper.  T.\,D.\,M.\ acknowledges support provided by NASA through
Chandra Postdoctoral Fellowship grant number PF8-10005 awarded by the
Chandra Science Center, which is operated by the Smithsonian
Astrophysical Observatory for NASA under contract NAS8-39073.

\newpage

\begin{figure}[t]
\centerline{\epsfysize=5.7in\epsffile{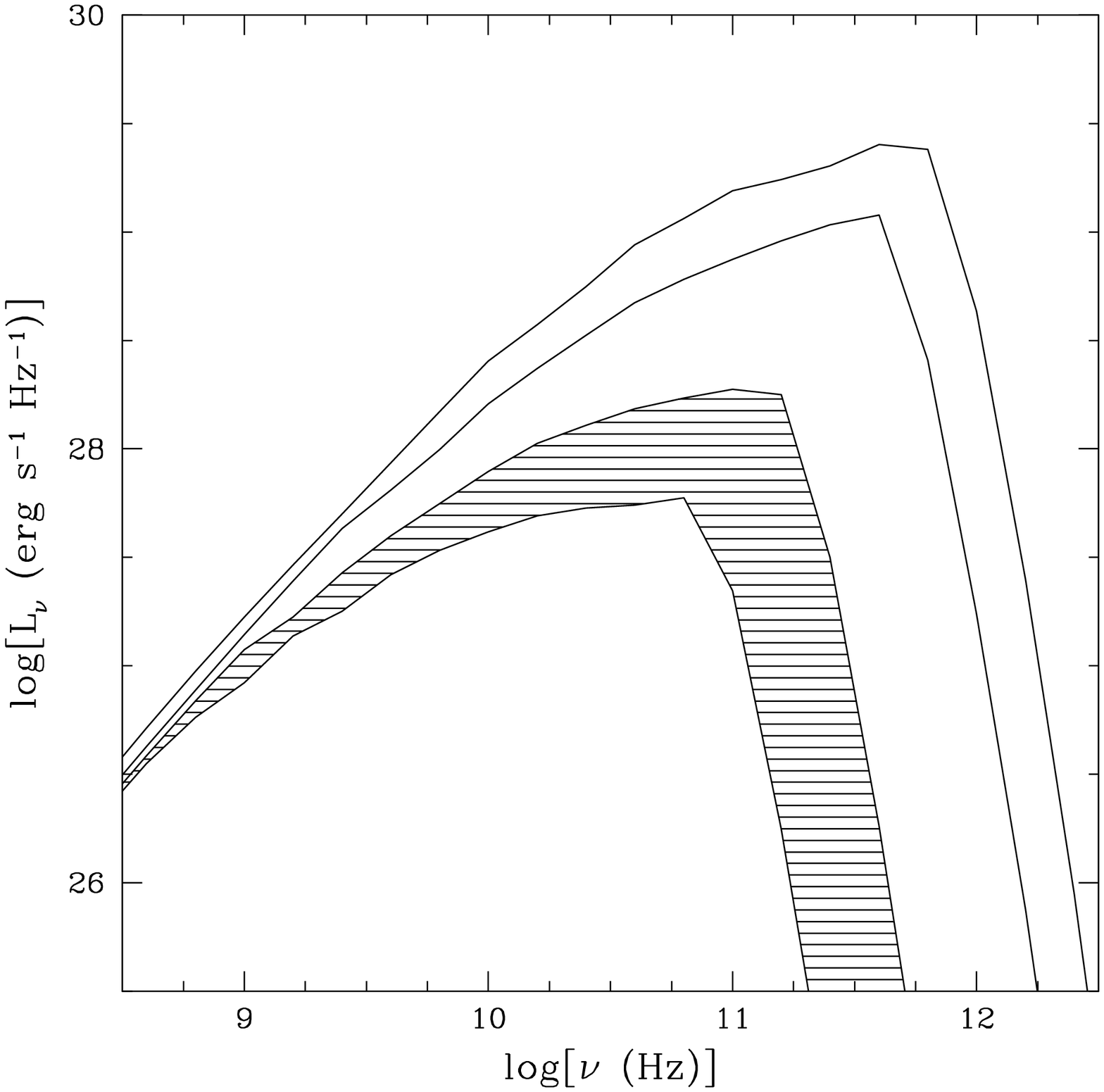}}
\caption{Synchrotron emission from low-radiative efficiency accretion
flows.  The various lines correspond to various amounts of mass loss
in the flows ($p$ = 0, 0.2, 0.4, 0.6 from top to bottom).  The
uppermost curve is the standard ADAF model with no outflows.  The
shaded region indicates the range of luminosities which best fit the
data in the cores of ellipticals observed so far (Di Matteo et
al. 2000b).}
\label{fig:1}
\end{figure}

\begin{figure}[t]
\centerline{\epsfysize=5.7in\epsffile{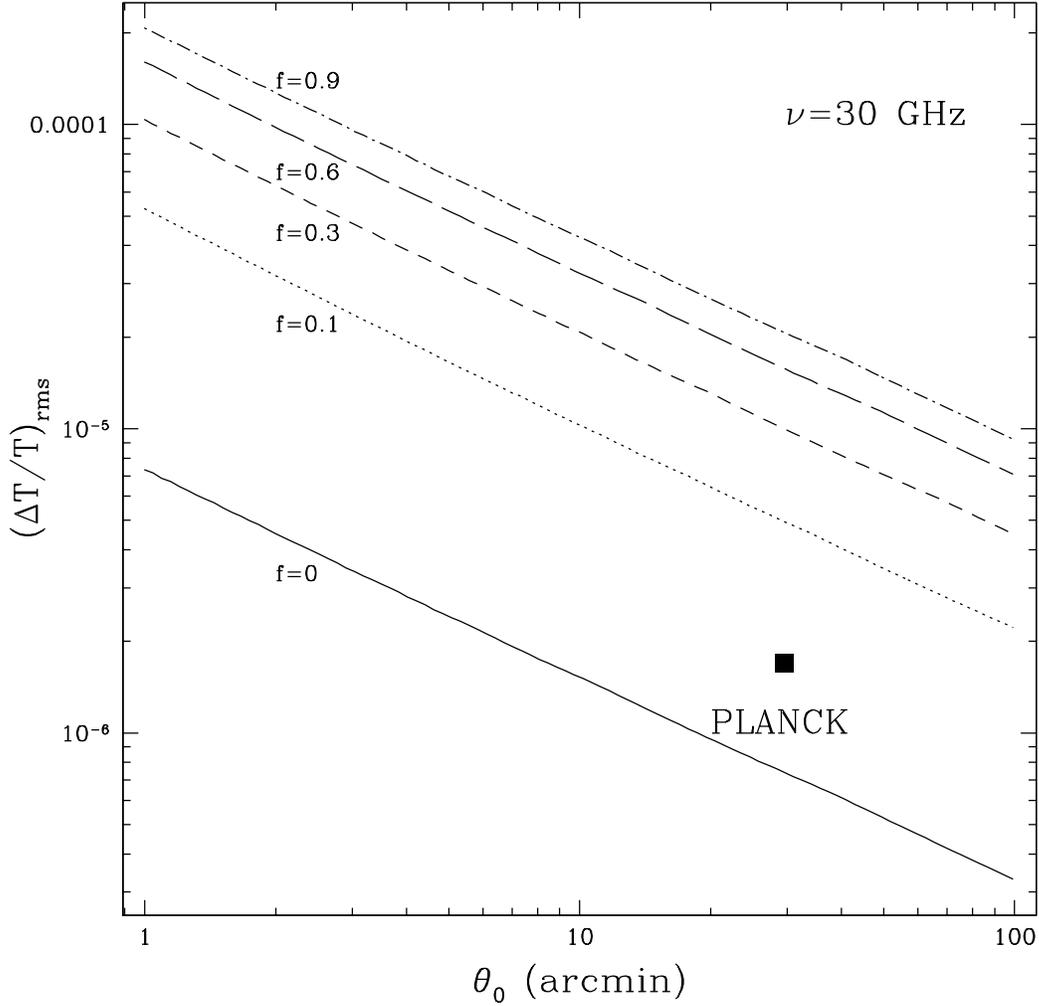}}
\caption{Potential contribution to the CMB anisotropy from accretion
flows in the cores of ellipticals as a function of the beamsize $\theta_0$
at $\nu = 30$ GHz.  The solid line shows the expected fluctuation
level if the luminosities of all sources were in the range
corresponding to the shaded region of Figure~1.  The other lines show
the fluctuations expected if a fraction $f$ of sources with a higher
luminosity (corresponding to the uppermost curve of Figure 1) were
mixed with these sources. The noise level indicated for Planck is the
average $\Delta T/T$ per pixel for 1$\sigma$ detection after two full
sky coverages.}
\label{fig:2}
\end{figure}

\begin{figure}[t]
\centerline{\epsfysize=5.7in\epsffile{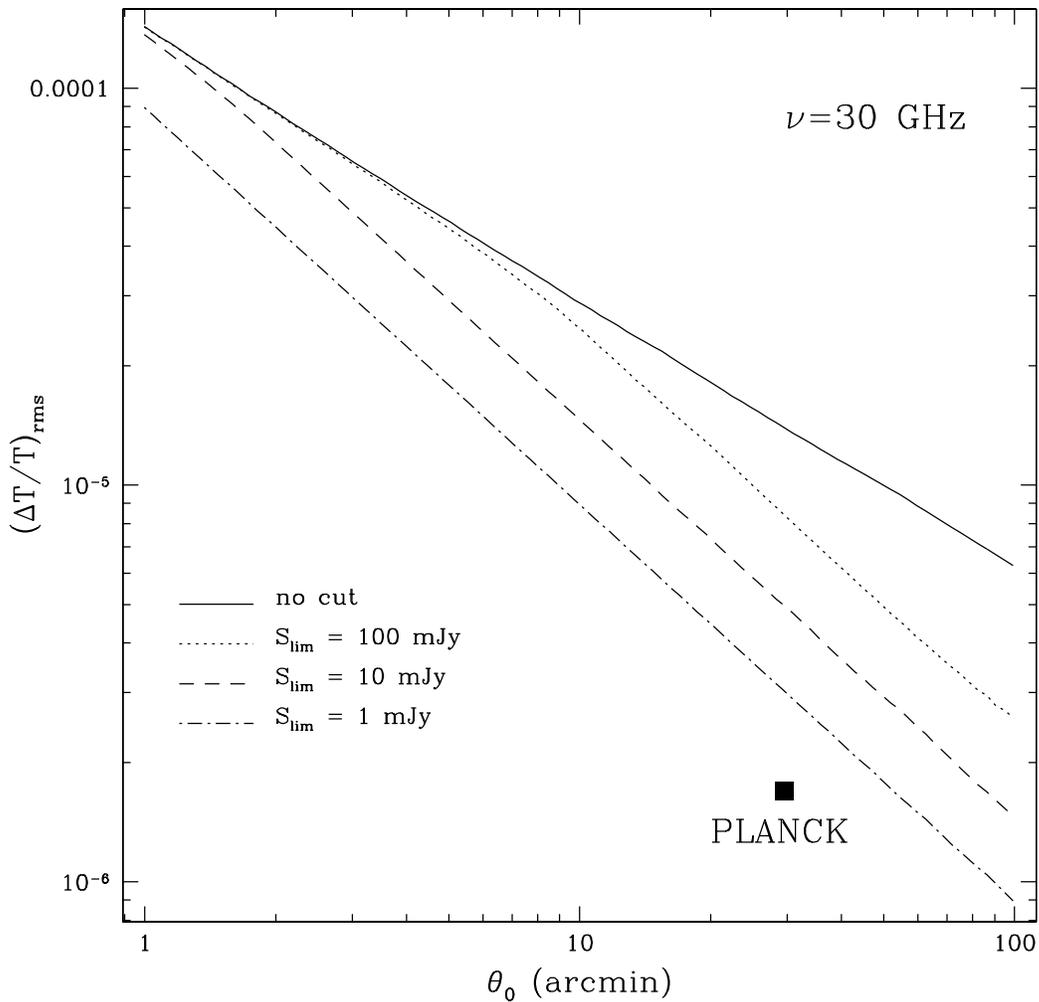}}
\caption{Fluctuations in the CMB anisotropy due to accretion flows  
in the cores of ellipticals, after removal of all sources above
a flux $S_{\rm lim}$. Here a fraction 
$f=0.5$ of high-luminosity sources has been assumed in the sample.}
\label{fig:3}
\end{figure}

\begin{figure}[t]
\centerline{\epsfysize=5.7in\epsffile{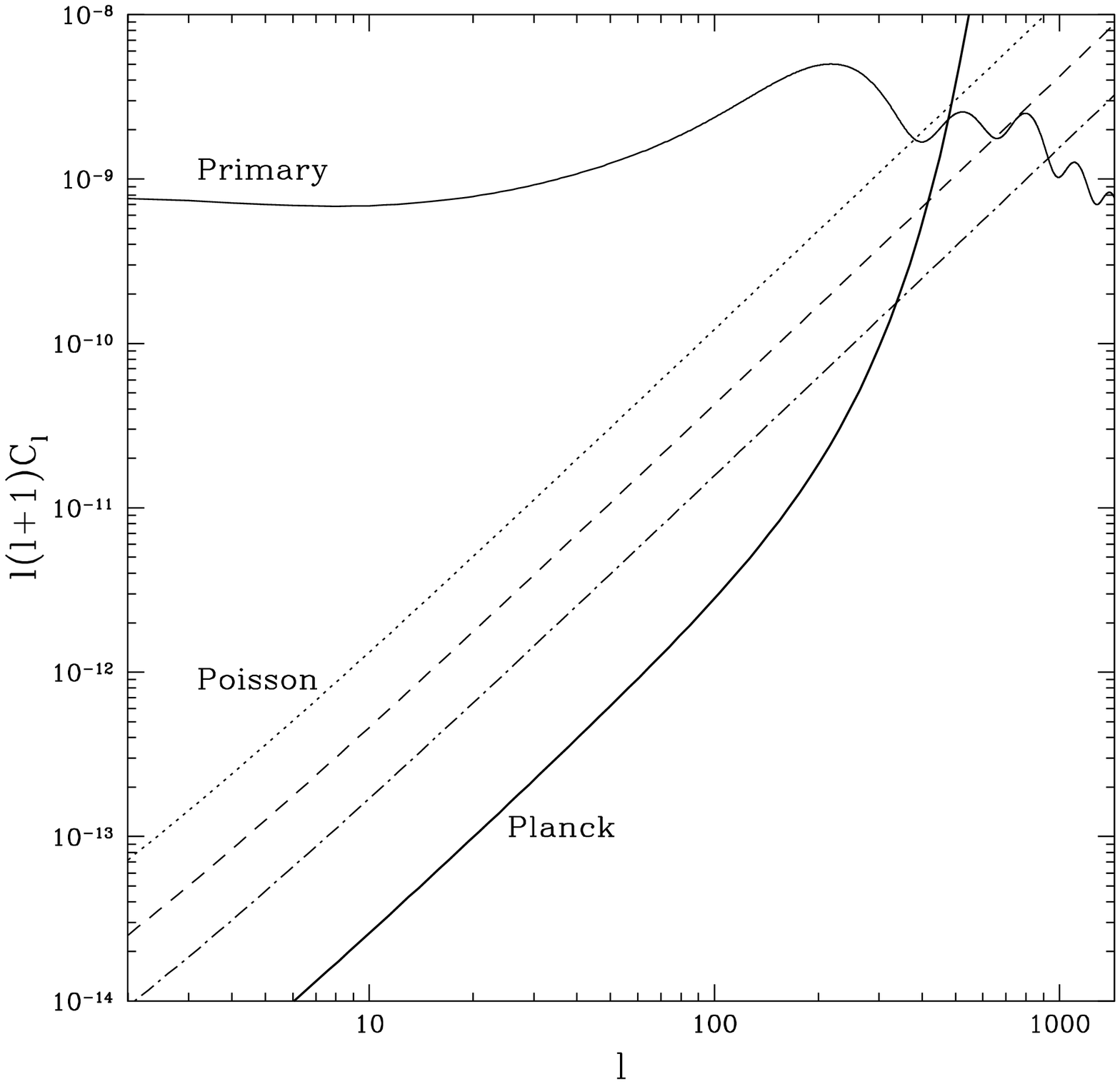}}
\caption{Poisson component of the angular power spectrum of our sources
for a range of flux cuts: $S_{\rm lim}=100$ mJy (dotted line), 
$S_{\rm lim}=10$ mJy (dashed line), $S_{\rm lim}=1$ mJy (dotted-dashed line).
This is shown for the case where a fraction $f=0.5$ of high-luminosity sources
is present in the sample. To compare with the level of primary anisotropy
expected, the prediction for a standard CDM spectrum is also shown, normalized
to COBE. The thick solid line is the expected contribution to the power spectrum 
from noise in the 30 GHz channel of the Planck LFI.}
\label{fig:4}
\end{figure}


\begin{references}

\reference{} Allen S.W., Di Matteo T., Fabian A.C., 2000, MNRAS, 311, 493

\reference{} Blandford, R. D. \& Begelman, M. C. 1999, \mnras, 303, L1

\reference{} Cavaliere, A. \& Setti, G. 1976, A\&A, 46, 81

\reference{} Condon, J. J. 1974, ApJ, 188, 279

\reference{} Cooray, A. R., Grego, L., Holzappel, W. L., Marshall, J.,
\& Carlstrom, J. E. 1998, ApJ, 115, 1388

\reference{} De Zotti, G., Granato, G. L., Silva, L., Maino, D., \& Danese, L.
astro-ph/9912282

\reference{} Di Matteo T., Allen S.W., 1999, ApJ, 527, L21

\reference{DM99b} Di Matteo, T., Fabian, A.\,C., Rees, M.\,J., Carilli, C.\,L.,
Ivison, R.\,J.\, 1999, MNRAS, 305, 492

\reference{DM99a} Di Matteo, T., Quataert, E., Allen, S.\,W., Narayan, R., Fabian
A.\,C., 2000a, MNRAS, 311, 507

\reference{} Di Matteo, T.,  Carilli, C.L., Fabian A.C., 2000b, ApJ, submitted

\reference{} Fabian A.~C., Rees M.~J., 1995, MNRAS, 277, L55

\reference{} Franceschini, A., Toffolatti, L., Danese, L. \& De Zotti. G.
1989, ApJ, 344, 35

\reference{} Franceschini, A., Vercellone, S., \& Fabian, A. C. 1998, MNRAS, 297, 817

\reference{} Gawiser, E., \& Smoot, G. F. 1997, ApJ, 480, L1

\reference{} Guerra, E. J., Haarsma, D. B., Partridge, R. B. 1998, AAS, 193, 4003

\reference{} Igumenshchev, I. V.  Abramowicz, M. A. 1999, \mnras, 303, 309 

\reference{} Kogut, A., et al. 1996, ApJ, 464, L5

\reference{} Loan, A. J., Wall, J. V. \& Lahav, O. 1997, MNRAS, 286, 994

\reference{} Mahadevan, R. 1997, ApJ, 477, 585

\reference{} Magorrian, J. et al. 1998, AJ, 115, 2285

\reference{} Mather, J. C., Fixsen, D. J., Shafer, R. A., Mosier, C., \&
Wilkinson, D. T. 1999, ApJ, 512, 511

\reference{} Mushotzky, R.F., Cowie, L.L., Barger, A.J., Arnaud A., 2000, Nature, in press

\reference{} Narayan, R. \& Yi, I. 1995, ApJ, 444, 231 

\reference{N99} Narayan, R., Mahadevan, R., Quataert, E.\ 1998,
Theory of Black Hole Accretion Disks, edited by Marek A. Abramowicz,
Gunnlaugur Bjornsson, and James E. Pringle. Cambridge University Press, p.148

\reference{} Philips, R. B. \& Mutel, R. L., 1982, A\&A, 106, 21

\reference{} Rees M.~J., Begelman M.~C., Blandford R.~D., Phinney E.~S.,
1982, Nature, 295, 17 

\reference{} Scheuer, P. A. G. 1957, {\em Proc. Cambridge Phil. Soc.}, 53, 764

\reference{} Scheuer, P. A. G. 1974, MNRAS, 166, 329

\reference{} Scott, D. \& White, M. 1999, A\&A, 346, 1

\reference{} Shaver, P. A. 1988, in 'IAU Symposium 130, Evolution of large Scale Structures
in the Universe', ed. J. Audouze, M.-C. pelletan, and A. Szalay
(Dordrecht: Reidel) 

\reference{} Slee O.B., Sadler E.M., Reynolds J.E., Ekers R.D., 1994, MNRAS, 269, 92

\reference{} Smoot, G. F., et al. 1992, ApJ, 396, L1

\reference{} Sokasian, A., Gawiser, E., \& Smoot, G. F. 1998, ApJ submitted, 
astro-ph/9811311 

\reference{} Stanghellini, C., O'Dea, C. P., Dallacasa, D., Baum, S. A.,
Fanti, R., \& Fanti, C. 1998, A\&AS, 131, 303

\reference{} Stone, J. M., Pringle, J. E.  Begelman, M. C. 1999, \mnras,
310, 1002

\reference{} Tegmark, M. \& Efstathiou, G. 1996, MNRAS, 281, 1297

\reference{} Toffolatti, L., Gomez, F. A., De Zotti, G., Mazzei, P.,
Franceschini, A., Danese, L. \& Burigana, C. 1999a, MNRAS, 297, 117

\reference{} Toffolatti, L., De Zotti, G., Argueso, F., \& Burigana, C.
astro-ph/9902343

\reference{} Wrobel J.~M., 1991, AJ, 101, 12

\reference{} Yi, I., \& Boughn, S. P. 1998, APJ, 499, 198

\end{references}
\end{document}